\documentclass[11pt]{article}
\newcommand{\ket}[1]{\left | #1 \right \rangle}
\newcommand{\bra}[1]{\left  \langle #1 \right |}
\newcommand{\eq}{\begin{equation}}
\newcommand{\en}{\end{equation}}
\newcommand{\braket}[2]{\left \langle #1 | #2 \right \rangle}
\newcommand{\overlap}[2]{| \left \langle #1 | #2 \right \rangle |}

\newtheorem{theorem}{Theorem}

\newcounter{count1}
\begin{document}
\begin{center}
{\Large\bf Towards a geometrical interpretation\\ of quantum information compression }\\
\bigskip
{\normalsize  Graeme Mitchison$^\S$ and Richard Jozsa$^\dagger$}\\
\bigskip

{\small\it
$^\S$MRC Laboratory of Molecular Biology,\\ Hills Road, Cambridge, CB2 2QH, U.K.\\[1mm]

$^\dagger$Department of Computer Science, University of Bristol,\\
Merchant Venturers Building, Bristol BS8 1UB U.K. }
\\[4mm]
\date{today}
\end{center}

\begin{abstract} Let $S$ be the von Neumann entropy of a finite
ensemble $\cal E$ of pure quantum states. We show that $S$ may be
naturally viewed as a function of a set of geometrical volumes in
Hilbert space defined by the states and that $S$ is monotonically
increasing in each of these variables. Since $S$ is the Schumacher
compression limit of $\cal E$, this monotonicity property suggests
a geometrical interpretation of the quantum redundancy involved in
the compression process. It provides clarification of previous
work in which it was shown that $S$ may be increased while
increasing the overlap of each pair of states in the ensemble. As
a byproduct, our mathematical techniques also provide a new
interpretation of the subentropy of $\cal E$.

\end{abstract}


\section{Introduction}\label{intro}

One of the most satisfying results of quantum information theory
is Schumacher's source coding theorem
\cite{Coding,Newproof,horod98,barnum,winter}, which says that a
message of length $K$ from a source of pure quantum states with
density matrix $\rho$ can be compresssed to $KS(\rho)$ qubits
(asymptotically for large $K$), where $S(\rho)$ is the von Neumann
entropy
\[
S(\rho)=-Tr(\rho \log \rho).
\]
Conceptually we can associate the possibility of compression with
the presence of a degree of redundancy in the source. Suppose
$\rho_1$, $\rho_2$ are the density matrices of two sources and \eq
S(\rho_1) \le S(\rho_2). \label{sinequal} \en Then the first
source can be compressed further than the second, so it has
greater redundancy. If the two sources both have $k$ states,
$\ket{\psi_i}$ and $\ket{\phi_i}$, say, with the same
probabilities $p_i$, then this increased redundancy does not lie
in the classical probabilities $p_i$ of emitting the states but in
the properties of the states themselves: it is true ``quantum
redundancy''. Intuitively, one expects that, if the states
$\ket{\psi_i}$ are more similar than the states $\ket{\phi_i}$,
then the quantum redundancy will be greater. So if the pairwise
overlaps of the $\ket{\psi_i}$ are larger than those of the
$\ket{\phi_i}$, i.e. \eq \overlap{\psi_i}{\psi_j} \ge
\overlap{\phi_i}{\phi_j} \mbox{ for all } i,j, \label{oinequal}
\en then $\rho_1$ should have more quantum redundancy than
$\rho_2$, and consequently (\ref{sinequal}) should hold. Jozsa and
Schlienz \cite{JS} showed this is indeed true for a source with
two states. However, they produced a counter-example consisting of
a set of three states, $\ket{\phi_i}$, in three dimensions and a
slight perturbation of them, $\ket{\psi_i}$, that has greater
overlaps but {\em larger} entropy. This phenomenon raises the
question of whether compression and quantum redundancy can be
understood in {\em geometrical} terms, i.e. in terms of the
geometry of the source's states in Hilbert space. This question
has also been recently raised in \cite{ibm}. In this paper we will
establish a connection between quantum redundancy and the volumes
in Hilbert space defined by the source states.

One reason the above example appears paradoxical is that, in real
three-dimensional geometry, the pairwise inner products of three
unit vectors determine the figure the vectors make, up to an
orthogonal transformation, and they also determine the entropy in
this real-valued setting. An analogous result holds in complex
geometry if the complex inner product is specified, but not if
only its absolute value, the overlap, is given. In fact, taking
into account the freedom in specifying phase for a quantum
mechanical state, it turns out that there are {\em four}
real-valued degrees of freedom in specifying three states in three
dimensions up to a unitary transformation (as will be shown
shortly). Specifying the three pairwise overlaps
$\overlap{\psi_1}{\psi_2}$, $\overlap{\psi_1}{\psi_3}$ and
$\overlap{\psi_2}{\psi_3}$ therefore leaves a further degree of
freedom that can be used to adjust the entropy.


What is this extra degree of freedom for three states in three
(complex) dimensions? Here it is useful to introduce the Gram
matrix $G$, with entries
$G_{ij}=\sqrt{p_ip_j}\braket{\psi_i}{\psi_j}$ and the matrix $A$
with entries $a_{ij}=\braket{\psi_i}{\psi_j}$. $G$ has the same
eigenvalues as the density matrix $\rho=\sum p_i
\ket{\psi_i}\bra{\psi_i}$ (see \cite{JS}) so the von Neumann
entropy \eq S(\rho)=-\sum x_i\log x_i \label{eigen} \en can be
computed from the eigenvalues $x_i$ of the Gram matrix
\[
G=\left(\matrix{ p_1a_{11} & \sqrt{p_1p_2}a_{12} &
\sqrt{p_1p_3}a_{13} \cr \sqrt{p_2p_1}a_{21} &  p_2a_{22} &
\sqrt{p_2p_3}a_{23} \cr \sqrt{p_3p_1}a_{31} & \sqrt{p_3p_2}a_{32}
& p_3a_{33} & \cr}\right).
\]
 $G$ has the characteristic equation
\eq x^3-s_1x^2+s_2x-s_3=0, \label{char3} \en where the $s_i$ are
the symmetric polynomial functions of the eigenvalues, with
$s_1=\sum x_i=1$, $s_2=\sum_{i<j} x_ix_j$ and $s_3=x_1x_2x_3$.
Expanding $\det(G-Ix)$ one finds that
\begin{equation}\label{ss}
s_2=\sum_{i<j} p_ip_j(1-|a_{ij}|^2) \hspace{5mm}\mbox{and}
\hspace{5mm} s_3=p_1p_2p_3 \det A.
\end{equation}
Since $A$ can be written as $A=BB^\dagger$, where $B$ is the
matrix
\[
\left(\matrix{ \psi_{11} & \psi_{12} & \psi_{13} \cr
\psi_{21} & \psi_{22} & \psi_{23} \cr
\psi_{31} & \psi_{32} & \psi_{33} \cr}\right)
\]
of coordinates $\psi_{ij}$ of the states $\ket{\psi_i}$ in any
orthonormal basis, it follows that $\det A=|\det B|^2$, and we can
regard $\det A$ as the squared modulus of the {\em complex volume}
spanned by the $\ket{\psi_i}$. Note also that $(1-|a_{ij}|^2)=\det
A_{i,j}$, where $A_{i,j}$ is the submatrix of $A$ obtained by
striking out rows and columns having labels not in the set $\{i,j
\}$. Thus we can likewise regard $1-|a_{ij}|^2$ as the squared
modulus of the complex volume spanned by $\ket{\psi_i}$ and
$\ket{\psi_j}$. We can write the terms of eq (\ref{ss}) that
depend on the states as \eq \alpha_{12}=\det A_{1,2}, \
\alpha_{13}=\det A_{1,3}, \ \alpha_{23}=\det A_{2,3} \mbox{ and }
\alpha_{123}=\det A \label{invar} \en and $s_2$ and $s_3$ then
appear as positive linear combinations of these volume variables.
Thus $S(\rho)$ is determined by the probabilities $p_i$ and the
four squared volumes in (\ref{invar}). If the states are perturbed
in such a way that three of these four parameters are fixed but
one of them varies, then the entropy changes as intuition would
dictate: increasing $A_{12}$ (i.e. decreasing the overlap between
states 1 and 2) increases the entropy (decreases the redundancy),
and increasing the volume (i.e.  spreading the states apart) also
increases the entropy. In this sense we can regard the four
parameters $\alpha_{12}, \alpha_{13}, \alpha_{23}, \alpha_{123}$,
as measures of quantum redundancy for the set of states. This
behaviour of the entropy follows from a theorem proved later,
which tells us that $\partial S/\partial s_i > 0$ for $i\geq 2$.
Consequently when none of the $p_i$ is zero (which we shall assume
throughout), $\partial S/
\partial \alpha_{ij}= p_ip_j\partial S/\partial s_2 > 0$ and
$\partial S/
\partial \alpha_{123}= p_1p_2p_3\partial S/\partial s_3 > 0$.

\section{k states in n dimensions}

We now investigate the situation for any number $k$ of states
which span a space of $n$ dimensions (so $k \ge n$). We can try to
imitate in $n$ dimensions the procedure that gave the variables in
(\ref{invar}). Let the states be $\ket{\psi_i}$ with probabilities
$p_i$ for $i=1,\ldots ,k$. As above we introduce the Gram matrix
with entries $G_{ij}=\sqrt{p_ip_j}\braket{\psi_i}{\psi_j}$ and the
$A$-matrix with entries $a_{ij}=\braket{\psi_i}{\psi_j}$, both
being $k\times k$ matrices. The eigenvalues of the Gram matrix are
those of the $n \times n$ density matrix $\rho$ padded out with
$k-n$ zeros \cite{JS}. Thus the characteristic equation $\det
(G-xI)=0$ for the Gram matrix for $k$ states in $n$ dimensions has
the form
\eq
(-1)^{k-n}x^{k-n}\sum_{i=0}^n (-1)^{i}s_{n-i}x^i=0,
\label{char}
\en
where $s_i$ is the $i$-th elementary symmetric
polynomial of the eigenvalues $x_1, \dots , x_n$ of the density
matrix, defined by
\begin{equation} s_0=1,\hspace{10mm} s_i=\sum_{u_1<\ldots <u_i}
x_{u_1}\ldots x_{u_i}\hspace{3mm}\mbox{for $i=1,\ldots ,n$}.
\end{equation}
The $i$th symmetric polynomial may be expressed as
\begin{equation}
\label{sidet}
s_i=\sum_{u_1<\ldots <u_i} p_{u_1} \ldots p_{u_i} \alpha_{u_1,
\ldots , u_i},
\end{equation}
where $\alpha_{u_1, \ldots , u_i}=\det A_{u_1, \ldots , u_i}$, and
$A_{u_1, \ldots , u_i}$ is the sub-matrix of the matrix $A$
obtained by striking out all rows and columns with labels not in
the set $\{u_1, \ldots , u_i\}$; i.e. $A_{u_1, \ldots , u_i}$ is
the $i\times i$ $A$-matrix constructed from the subset $\{
\ket{\psi_{u_1}}, \ldots ,\ket{\psi_{u_i}} \}$ of the states. This
follows because $s_i$ is the coefficient of $(-x)^{k-i}$ in
$\det(G-xI)$, which is obtained by picking $k-i$ elements on the
diagonal, corresponding to rows $v_1, \ldots, v_{k-i}$ say, and
for each such choice of $v$'s, constructing the determinant of the
sub-matrix $G_{u_1,\ldots ,u_i}$ of $G$ where $u_1, \ldots , u_i$
is the complementary set to the $v$'s. Since the Gram matrix $G$
is related to the corresponding $A$-matrix by $G=QAQ$, where $Q=
{\rm diag}(\sqrt{p_1}, \ldots ,\sqrt{p_k})$, we get $\det
G_{u_1,\ldots ,u_i}=p_{u_1}\ldots p_{u_i}\det A_{u_1,\ldots
,u_i}$. Eq (\ref{sidet}) is then obtained by summing over all sets
of $u$'s.

Note that, though the individual entries in $A_{u_1, \ldots , u_i}$
are dependent on a choice of phase for the states $\ket{\psi_i}$,
$\alpha_{u_1, \ldots , u_i}$ is invariant under phase
choices. Furthermore, $\alpha_{u_1, \ldots , u_i}$ is real, since
$A_{u_1, \ldots , u_i}$ is Hermitian. Thus $\alpha_{u_1, \ldots ,
u_i}$ is a real-valued unitary invariant, and the complete set of all
$\alpha_{u_1, \ldots , u_i}$ for all sets $u_1, \ldots , u_i$ can be
regarded as the analogues of the invariants (\ref{invar}) in the case
$k=n=3$.  Also, as in the case of $k=n=3$, $A_{u_1,\ldots
,u_i}=B_{u_1,\ldots ,u_i}B^\dagger_{u_1,\ldots ,u_i}$, where
$B_{u_1,\ldots ,u_i}$ is the $i\times i$ matrix whose rows are the
components of the $i$ states $\ket{\psi_{u_1}},\ldots
,\ket{\psi_{u_i}}$ (expanded in any choice of orthonormal basis in the
span of these $i$ states).  Thus $\alpha_{u_1,\ldots ,u_i}$ may be
identified as the squared modulus of the complex volume determined by
$\ket{\psi_{u_1}},\ldots ,\ket{\psi_{u_i}}$.

Let $x_1,\ldots ,x_n$ be any probability distribution and let
$s_i=s_i(x_1,\ldots ,x_n)$ for $i=1,\ldots ,n$ be the corresponding
symmetric polynomials.  For any function $f(x_1, \ldots ,x_n)$
(e.g. the entropy $S$) we consider a change of variables from the
$x_i$'s to the $s_j$'s. Note that the probability condition $\sum
x_i=1$ corresponds to $s_1=1$, and lifting this condition we get $n$
variables in each case. Then the Jacobian is readily seen to be
$\prod_{i<j} (x_i-x_j)$, so the change of variables is valid if the
$x_i$'s are all different. For simplicity we will work within this
restriction but expect that our results will have suitable (finite)
limiting behaviour for coincident values $x_i\rightarrow
x_j$. Furthermore we will be interested primarily in partial derivatives
$\partial f/\partial s_i$ for $i\geq 2$ (which have $s_1$ held
constant), so our results will also remain valid if we impose the
probability constraint $s_1=1={\rm constant}$ at the start.

We have the following fundamental property of the entropy:
\begin{theorem}\label{theorem1} If $S=-\sum x_i \log x_i$ is
viewed as a function of the symmetric polynomials $s_1, \ldots
,s_n$ then $\partial S/\partial s_q >0$ for $q=2,\ldots ,n$.
\end{theorem}

Two proofs of this theorem are given in the appendix.

\section{Geometrical interpretation of quantum
redundancy}\label{sect3}

Eq (\ref{sidet}) gives an expression for the symmetric polynomials
$s_i$ that is canonically determined by the state set $\{
\ket{\psi_1},\ldots ,\ket{\psi_k}\}$ and probabilities $p_1,
\ldots ,p_k$. Thus from $S=S(s_1,\ldots ,s_n)$ (with $s_1=1$) we
can view the von Neumann entropy of the source in a natural way as
being a function of the probabilities and all the squared volumes
$\alpha_{i_1i_2}, \ldots ,\alpha_{i_1 i_2 \ldots i_k}$. By theorem
\ref{theorem1}, $\partial S/\partial s_q>0$ for $q\geq 2$ and by eq
(\ref{sidet}), each $s_q$ is a positive linear combination of the
$\alpha$-variables, so we conclude that $\partial S/\partial \alpha
>0$ for each squared volume variable $\alpha$.

This suggests a geometrical interpretation of the quantum
redundancy in the ensemble of quantum states. If the probabilities
are held fixed and the states are deformed then the change in
entropy can be seen as an accumulation of monotonic effects
arising from the changes induced in each of the squared volumes
$\alpha$. Since $\partial S/\partial \alpha >0$ the set of these
squared volumes can be regarded as a geometric measure of the
quantum redundancy associated to a set of states alone.

Note however that, for $k>3$, there are more $\alpha$'s than
degrees of freedom needed to fix $k$ states up to overall unitary
equivalence. Indeed, let $\nu(k,n)$ denote the number of degrees
of freedom in specifying $k$ states in $n$ dimensions up to
unitary transformation. To specify $k$ states requires $k(2n-2)$
real parameters (as each state is defined only up to overall
phase). The unitary group $U(n)$ has $n^2$ parameters, but because
of the overall phase freedom in each state, $U$ and $e^{ix}U$ have
the same action for any $x$. Thus, unitary action on the states
eliminates $n^2-1$ parameters from the $k(2n-2)$, giving
$\nu(k,n)=k(2n-2)-(n^2-1)$.

The following table shows $\nu(k,n)$ for small values of $k$ and
$n$ and $k \ge n$. The bracketed numbers are the total number
$\tau(k,n)$ of terms $\alpha_{u_1, \ldots , u_i}$, ignoring those
with $i>n$, which are zero since more than $n$ states must be
linearly dependent and therefore have zero determinant. The
numbers in brackets are therefore $\tau(k,n)=\sum_{i=2}^n {k
\choose i}$.

\[
\begin{array}{ccccc} \multicolumn{5}{c}{\rm Table} \cr k  & n=2 & n=3 & n=4 & n=5
\cr 2 & 1 \ (1)   &     &       &    \cr 3  & 3 \ (3) & 4 \ (4)  &
& \cr 4  &  5 \ (6)   &   8  \ (10) & 9 \ (11) &   \cr 5  & 7  \
(10) & 12 \ (20) &   15 \ (25)  & 16 \ (26) \end{array}
\]
For $k \le 3$ the two sets of numbers agree, so the $\alpha$'s can
be used to parametrize the sets of states. For $k>3$ there are
always too many $\alpha$'s. Thus viewing the entropy $S$ as a
function of the $\tau(k,n)$ $\alpha$'s (for fixed probabilities)
amounts to a non-trivial extension of $S$ to a larger space of
variables: not every $\tau(k,n)$-tuple of $\alpha$-values is
geometrically realisable by an ensemble of $k$ states in $n$
dimensions, and when an actual ensemble is deformed, the
$\alpha$-variables are constrained to lie on a surface of
dimension $\nu(k,n)$ in the ambient space of dimension
$\tau(k,n)$. But the virtue of this non-physical extension of the
number of parameters is that we are able to attribute
compressibility of the source to geometrical constructs, viz. the
$\alpha$'s. In any deformation of actual states, each $\alpha$
varies positively or negatively and the compressibility varies by
a corresponding accumulation of monotonic positive and negative
effects.

\section{Minimal sets of monotonic parameters?}

Since $\tau(k,n)>\nu(k,n)$ for $k>3$, it is interesting to ask
whether we can find some alternative set of parameters, $\beta_1,
\ldots , \beta_{\nu(k,n)}$ which is in 1-1 correspondence with the
set of $k$ states and has some of the desirable properties
possessed by the $\alpha$'s in the case $k \le 3$. In particular,
to make a connection with the phenomenon of compression, we would
like the monotonicity property $\partial S/
\partial \beta_i>0$ to hold for any choice of probabilities $p_i$.
Intuitively, this means that we can regard the $\beta$'s as
measures of quantum entropy.

Consider the case $k=n=4$. Here the set of states is 9 dimensional,
whereas there are 11 $\alpha$'s. Can we perhaps keep some of the
$\alpha$'s, taking, say, the overlap-related terms $\alpha_{ij}$ as
$\beta_1, \ldots \beta_6$, and adding a further 3 $\beta$'s?  (For
instance, one might consider adding the 3 distances between subspaces
generated by disjoint pairs of the four states, i.e. $d(12,34)$,
$d(13,24)$ and $d(14,23)$, where $d(ij,kl)$ is some measure of the
distance between the subspace spanned by $\ket{\psi_i}$,
$\ket{\psi_j}$ and that spanned by $\ket{\psi_k}$, $\ket{\psi_l}$).
However, it turns out that {\em any} set of parameters that includes
the 6 terms $\alpha_{ij}$ cannot have the desired monotonicity
property. To see this, write $r_{ij}=|a_{ij}|$ and define
$u=\arg(a_{12}a_{23}a_{31})$, $v=\arg (a_{14}a_{21}a_{42})$, and
$w=\arg(a_{13}a_{34}a_{41})$. Then we can write all the $\alpha$'s in
terms of the 6 $r_{ij}$'s and $u$, $v$ and $w$, giving 9 parameters in
all. For instance,
$\alpha_{123}=1-r_{12}^2-r_{23}^2-r_{31}^2+2r_{12}r_{23}r_{31}\cos u$.

Now pick one of the $\beta_i$ for $i>6$, and call it $x$. Taking the
partial derivative with respect to $x$, the fact that
$\beta_1,...\beta_6$ are constant implies $\partial s_2/\partial x=0$
(since by eq (\ref{sidet}) $s_2$ is a function only of $\beta_1,\ldots
,\beta_6$). Furthermore, if we choose a set of states with $r_{ij}>0$
for all $i,j$ and $u=v=w=\pi/2$ we find
\begin{eqnarray*}
\partial s_3/ \partial x&=&-2(p_1p_2p_3)(r_{12}r_{23}r_{31})
u_x-2(p_1p_2p_4)(r_{14}r_{21}r_{42}) v_x\\
&&-2(p_1p_3p_4)( r_{13}r_{34}r_{41}) w_x+2(p_2p_3p_4)
( r_{23}r_{34}r_{42}) (u_x+v_x+w_x).
\end{eqnarray*}
Suppose the first three terms in the above expression for $\partial
s_3/\partial x$ are positive. Since $\beta_1, \ldots ,\beta_6$ are
real and positive, this means $u_x<0$, $v_x<0$, $w_x<0$. So the fourth
term is negative. By taking $p_1$ small enough, we can ensure that
$\partial s_3/\partial x<0$ and also that $\partial s_4/\partial x$ is
sufficiently small to ensure that the term with $q=3$ dominates the
sum $\partial S/\partial x = \sum (\partial S/ \partial s_q) (\partial
s_q/\partial x)$. So $\partial S/\partial x <0$ and monotonicity
fails. Suppose on the other hand that at least one of the first three
terms is negative, the term with $u_x$ say, so $u_x>0$. Then taking
$p_4$ small enough leads to the same conclusion.

This result suggests that it may be difficult to construct a
minimal set of $\nu(k,n)$ monotonic parameters which also has a
simple geometrical interpretation, thus further underlining the
benefits of considering the non-minimal parameter set in section
\ref{sect3}.

\section{A remark on subentropy}

It is curious that our proof of the theorem that the entropy is an
increasing function of the symmetric functions $s_q$ (Appendix)
depends upon an algebraic expression closely related to a quantity
called the subentropy. Given any density matrix $\rho$, the
subentropy $Q(\rho)$ \cite{JRW} is the greatest lower bound on the
accessible information of any ensemble of pure states
$\ket{\phi_1}, \ldots , \ket{\phi_m}$ with probabilities $p_1,
\ldots, p_m$, for which $\rho=\sum p_i\ket{\phi_i}\bra{\phi_i}$.
In terms of the eigenvalues $x_1,\ldots ,x_n$ of $\rho$ (or indeed
for any classical probability distribution) the subentropy can be
written
\[
Q(\rho)=-\sum_k \frac{x_k^n\log x_k}{\prod_{i \ne k}(x_k-x_i)}
\]
which is closely related to our eq (\ref{small}); i.e. we would
wish to put $q=0$ in that formula!

Looking at the derivation of eq (\ref{small}) (c.f. especially eq
(\ref{derchar})) we see that the value $q=0$ would correspond to a
coefficient, $c_0$ say, of $x^n$ in the characteristic equation
and then $\partial S/\partial c_0$ would be essentially the
subentropy. Thus let us divide through eq (\ref{chareq}) by $s_1$,
introducing new variables
\[
t_1=\frac{1}{s_1},\hspace{5mm} t_2=\frac{s_2}{s_1},\hspace{3mm} \ldots
\hspace{3mm}, t_n=\frac{s_n}{s_1} \] and look at
\begin{equation}\label{ptw}
\begin{array}{rcl} \tilde{p}(x)=\frac{1}{s_1}p(x) & = & t_1x^n -
x^{n-1}+ \ldots + (-1)^q t_q x^{n-q} + \ldots + (-1)^n t_n \\
 & = & \frac{1}{s_1}(x-x_1)\ldots (x-x_n).
\end{array}
\end{equation}
Viewing this equation as defining $x_i=x_i(t_1,\ldots , t_n)$ and
carrying out an implicit differentiation with respect to $t_1$ we
get
\[
\frac{\partial x_k}{\partial t_1} = -\frac{x_k^n
s_1}{\prod_{k\neq i} (x_k-x_i)}
\]
and
\[
\frac{\partial
S}{\partial t_1}= s_1 \sum_k \frac{(1+\log x_k)x_k^n}{\prod_{k\neq
i} (x_k-x_i)}.
\]
Finally using the identity (valid for any $x_1,\ldots ,x_n$)
\[
\sum_{k=1}^n \frac{x_k^n}{\prod_{k\neq i}
(x_k-x_i)} = x_1+ \ldots +x_n = s_1 \] we get \[ \frac{\partial
S}{\partial t_1}= s_1(s_1-Q).
\]
If $x_1, \ldots ,x_n$ is a
probability distribution, so $s_1=1$, then we get $\partial
S/\partial t_1=1-Q$. Thus we have proved:
\begin{theorem}\label{theorem2} Let $S=-\sum x_i \log x_i$ be the
Shannon entropy function defined on $\{ (x_1, \ldots ,x_n): x_i
>0 \mbox{ all $i$} \} $ (i.e. we lift the probability condition
$\sum x_i =1$). If $S$ is viewed as a function of $t_1=1/s_1$,
$t_2=s_2/s_1$, $\ldots$ , $t_n=s_n/s_1$ then at points with
$s_1=\sum x_i =1$ the subentropy is given by $Q(x_1, \ldots
,x_n)=1-\partial S/\partial t_1$.
\end{theorem}
Note that the above mathematical characterisation of subentropy
applies equally well within {\em classical} information theory (as
it is a derivative property of the Shannon entropy function), in
contrast to all previous work on subentropy \cite{JRW,NW} where it
relates only to {\em quantum} mechanical considerations
(especially the theory of information gain from quantum
measurements).

Finally we also note that there are other possible ways of getting
a nontrivial coefficient of $x^n$ in eq (\ref{chareq}). For
example instead of dividing through by $s_1$ we could divide
through by $s_n$ and introduce the variables \[
r_q=\frac{s_{n-q}}{s_n} = \mbox{$q^{\rm th}$ symmetric polynomial
of $1/x_1, \ldots ,1/x_n$}. \] We then get the equation
\[ \begin{array}{rcl} \frac{1}{s_n}p(x) & = &  r_n x^n + \ldots
+ (-1)^q r_{n-q}x^{n-q} + \ldots +(-1)^n \\ & = & (x/x_1-1)\ldots
(x/x_n-1) =0 \end{array} \] leading to an alternative
characterisation of subentropy $Q$ as \[ \frac{\partial
S}{\partial r_n}=s_n(s_1-Q) \] and the condition for $x_1,\ldots
,x_n$ to be a probability distribution is now $s_1=r_{n-1}/r_n=1$,
i.e. $r_n=r_{n-1}$.

\section{Discussion}

The problem we have addressed in this paper is whether there are
real-valued functions $\alpha_q$ of $k$ states in $n$ dimensions
that together characterize those states up to a unitary
transformation and are also ``measures of quantum redundancy'' in
the sense that $\partial S/\partial \alpha_q
>0$ for each $\alpha_q$. In other words, increasing one $\alpha$ while
holding the others fixed increases the entropy $S$ and hence
reduces the redundancy of the set of states. We would also like
the $\alpha_q$ to have an interpretation in terms of the Hilbert
space geometry of the states.

We use the term ``quantum redundancy'' here because we require
that $\partial S/\partial \alpha_q>0$ holds for any choice of
probabilities $p_i$ of the states $\ket{\psi_i}$, and the $p_i$
can be thought of as embodying the classical aspect of redundancy. Of
course, one might ask whether there are joint functions of the
states and their probabilities that characterize the entropy, and
one example of this is the ``perimeter'' considered recently by
Hartley and Vedral \cite{HV}. The question then is why one such
function should be preferred to another; after all, the symmetric
functions $s_q$ trivially determine the entropy via the
characteristic equation (\ref{char}). The functions in \cite{HV}
are motivated by the possibility of experimental measurement
whereas our considerations are motivated by a desire to
geometrically characterise a notion of quantum redundancy in
quantum information compression.

Our main conclusion is that there is a natural set of measures (in
our sense) for sets of two or three states, but for four or more
states the corresponding parameters -- the determinants of
square submatrices of the matrix $(\braket{\psi_i}{\psi_j})$ --
outnumber the degrees of freedom in the sets of states, and the
more obvious ways of carrying over the results from two or three
states fail. Nevertheless the simple geometrical
interpretation of these parameters, in terms of Hilbert space
volumes defined by the states, makes it appealing to consider an
extension of the entropy function to the full space of these
variables, and the entropy of any physical ensemble of states then
appears as a special case satisfying some extra algebraic
constraint equations.\\[4mm]
{\Large\bf Acknowledgements}\\ RJ is supported by the UK
Engineering and Physical Sciences Research Council.

\section{Appendix: proof of theorem \ref{theorem1}}

We give here two proof of the theorem that $\partial S/\partial
s_q > 0$ for $2 \le q \le n$.\\[2mm]
{\bf First proof}\\ We will prove a slightly stronger result,
giving a positive lower bound for $\partial S/\partial s_q$ (see
eq (\ref{bound})).

Recall that
\[
S=-\sum_{i=1}^n x_i \log x_i,
\]
where the $x_i$ are roots of the characteristic equation (\ref{char})
\begin{equation}\label{chareq}
p(x)=\sum_{i=0}^n(-1)^{n-i}s_{n-i}x^i=0. \end{equation} Viewing
this equation as implicitly defining $x_i=x_i(s_1, \ldots ,s_n)$
and differentiating it with respect to $s_q$, we get
\begin{equation}\label{derchar}
\frac{\partial x_k}{\partial s_q}\left[\sum_{i=o}^n
(-1)^{n-i}s_{n-i}ix_k^{i-1}\right]+(-1)^qx_k^{n-q}=0.
\end{equation}
Since the expression in square brackets is the derivative of
$\prod(x-x_i)$ with $x$ set to $x_k$, we have
\[
\frac{\partial x_k}{\partial s_q}=\frac{(-1)^{q+1}x_k^{n-q}}{\prod_{i \ne k}(x_k-x_i)}.
\]
>From the chain rule, for any function $f$ of the $s_q$, \eq
\frac{\partial f}{\partial s_q} = \sum_{k=1}^n \frac{\partial
f}{\partial x_k}\frac{\partial x_k}{\partial s_q}. \label{chain}
\en Taking $f=s_1$ we get \eq \sum_{k=1}^n
\frac{x_k^{n-q}}{\prod_{i \ne k}(x_k-x_i)}=0 \mbox{ if } 2 \le q
\le n. \label{power} \en Then, taking $f=S=-\sum x_k \log x_k$
gives \eq \frac{\partial S}{\partial
s_q}=(-1)^q\sum_{k=1}^n\frac{x_k^{n-q}(1+\log x_k)}{\prod_{i \ne
k}(x_k-x_i)}, \en which in view of eq (\ref{power}) implies \eq
\frac{\partial S}{\partial s_q}=(-1)^q\sum_k \frac{x_k^{n-q}\log
x_k}{\prod_{i \ne k}(x_k-x_i)}, \mbox{ for } 2 \le q \le n.
\label{small} \en Define
\[
W_q(a)=(-1)^q \sum_k \frac{(x_k+a)^{n-q}\log(x_k+a)}{\prod_{i \ne k}(x_k-x_i)}
\]
so $W_q(0)=\partial S/\partial s_q$. Rewriting $W_q(a)$ as
\[
W_q(a)=(-1)^q a^{n-q} \sum_k \frac{(1+x_k/a)^{n-q}\log(1+x_k/a)}{\prod_{i \ne k}(x_k-x_i)},
\]
eq (\ref{power}) allows us to approximate $W_q(a)$ for large $a$ by the term in
$x_k^{n-1}$ in the expansion of $(1+x_k/a)^{n-q}\log(1+x_k/a)$, giving
\begin{eqnarray*}
W_q(a) &\simeq& (-1)^q a^{1-q} \times \left[\mbox{ coefficient of } x^{n-1} \mbox{ in } (1+x)^{n-q} \log(1+x)\right]\\
&=&a^{1-q}\int_0^1 y^{q-2}(1-y)^{n-q} dy\\
&\rightarrow 0& \mbox{ as } a \rightarrow \infty \mbox{ for } 2\le q \le n.
\end{eqnarray*}
Using eq (\ref{power}) again we find
\eq
\frac{\partial W_q(a)}{\partial a}=-(n-q)W_{q+1}(a),
\label{induct}
\en
for $2 \le q <n$, and applying eq (\ref{power}) to the set $a, x_1, \ldots, x_n$
\eq
\frac{\partial W_n(a)}{\partial a}=-1/\prod(a+x_k).
\label{ncase}
\en
Using $\prod_{k=1}^n(a+x_k) \le (a+1/n)^n$ in the preceding equation,
\[
W_n(x) \ge \int_x^\infty
\frac{da}{(a+1/n)^n}=\frac{1}{(x+1/n)^{n-1}(n-1)}
\]
so $\partial S/\partial s_n=W_n(0)\ge n^{n-1}/(n-1)$.

Eq (\ref{induct}) for $q=n-1$ then implies
\begin{eqnarray*}
W_{n-1}(x)&=& \int_x^\infty W_n(a)da \ge \int_x^\infty \frac{da}{(n-1)(a+1/n)^{n-1}}\\
&=&\frac{1}{(x+1/n)^{n-2}(n-1)(n-2)}.
\end{eqnarray*}
so $\partial S/\partial s_{n-1}=W_n(0) \ge n^{n-2}/(n-1)(n-2)$. And continuing this way we find
\eq
\partial S/\partial s_{n-q+1} \ge \frac {n^{n-q}}{q {n-1 \choose q}}.
\label{bound} \en So all the partial derivatives are bounded away
from zero, which proves the theorem.\\[2mm]
{\bf Second proof}\\  A second proof involves using a theorem from
numerical analysis -- the so-called Hermite-Gennochi
theorem \cite{atkinson} -- to replace the explicit derivation above
from eq (\ref{small}) onwards. (This theorem was also used in
\cite{JRW}, end of appendix A). Thus we begin as above, deriving
the expression in eq (\ref{small}) for $\partial S/\partial s_q$.

Now if $f(x)$ is any function whose values are known only at $n$
points $x_1,\ldots , x_n$ then there is a unique polynomial of
degree $n-1$, the Lagrange interpolating polynomial, that agrees
with the function at these points. The coefficient of $x^{n-1}$ is
called the Newton divided difference of $f$ and has standard
explicit formula $\sum_i f(x_i)/\prod_{k\neq i} (x_k-x_i)$. Thus
eq (\ref{small}) states that $\partial S/\partial s_q$ is the
Newton divided difference for the function $f(x)=(-1)^q
x^{n-q}\log x$. Now the Hermite-Gennochi theorem
asserts that the Newton divided difference is also given by the
integral over the probability simplex $\{ (p_1,\ldots ,p_n):
p_i\geq 0, \sum_i p_i=1 \}$ of $f^{(n-1)}(p_1x_1+\ldots p_nx_n)$
where $f^{(n-1)}$ is the $(n-1)^{\rm th}$ derivative of $f$.
Taking $f$ to be $(-1)^q x^{n-q}\log x$ it is straightforward to
check that $f^{(n-1)}(x)>0$ for all $0<x<1$. Hence the integral
over the probability simplex is positive and we get $\partial
S/\partial s_q >0$.


\begin{thebibliography}{}

\bibitem{Coding}
Schumacher, B. (1995) Quantum coding. Phys. Rev. A, {\bf 51}, 2738-2747.

\bibitem{Newproof} Jozsa, R. \& Schumacher, B. (1994) A new proof of the
quantum noiseless coding theorem, J. Mod. Opt. {\bf 41}, 2343-9.

\bibitem{horod98} Horodecki, M. (1998) Limits for compression of
quantum information carried by ensembles of mixed states, Phys.
Rev. A {\bf 57}, 3364.

\bibitem{barnum} Barnum, H., Fuchs, C., Jozsa, R. and Schumacher,
B. (1996) General fidelity limit for quantum channels, Phys. Rev.
 A {\bf 54}, 4707.

\bibitem{winter} Winter, A. (1999) Coding theorems of quantum
information theory. PhD thesis, ch. 1, University of Bielefeld,
Fakult\"{a}t f\"{u}r Mathematik. (Available at
http://xxx.lanl.gov/abs/quant-ph/9907077.)

\bibitem{JS}
Jozsa, R. and Schlienz, J. (1999) Distinguishability of states and
von Neumann entropy, Phys Rev A {\bf 62}, 012301-1. to 01203-11.

\bibitem{ibm} Jozsa, R. (2003) Illustrating the concept of quantum
information, IBM J. Res. Dev. (to appear). Available at
http://xxx.lanl.gov/abs/quant-ph/0305114.

\bibitem{JRW}
Jozsa, R., Robb, D. and Wootters, W.K. (1994) Lower bound for
accessible information in quantum mechanics. Phys Rev A {\bf 49},
668-677.

\bibitem{NW} Nichols, S. and Wootters, W.K. (2003) Between entropy
and subentropy, Quant. Inform. and Comp. {\bf 3}, 1-14.

\bibitem{atkinson} Atkinson, K.E. (1978) An introduction to
numerical analysis, (Wiley, New York) pp. 107-123.

\bibitem{HV}
Hartley, J. and Vedral, V. (2003) Entropy as a function of
geometric phase. Preprint available at
http://xxx.lanl.gov/abs/quant-ph/0309088.

\end{thebibliography}
\end{document}